\documentclass[aps,prd,showpacs,amssymb,nofootinbib,superscriptaddress,twocolumn]{revtex4-1}
\usepackage{graphicx,epstopdf}  % needed for figures
\usepackage{dcolumn}   % needed for some tables
\usepackage{bm,bbm,braket,soul,color}        % for math
\usepackage{amssymb,MnSymbol}   % for math
\usepackage{hyperref}

\usepackage{cancel} % strikethrough text

\DeclareMathOperator{\diag}{diag}
\DeclareMathOperator{\tr}{tr}
\DeclareMathOperator{\arctanh}{arctanh}

\begin{document}

\title{Generating entanglement between two-dimensional cavities in uniform acceleration}

\author{Bartosz Regula}
\email{pmxbr@nottingham.ac.uk}
\affiliation{School of Mathematical Sciences, University of Nottingham, University Park,
Nottingham NG7 2RD, United Kingdom}
\author{Antony R. Lee}
\email{pmaal1@nottingham.ac.uk}
\affiliation{School of Mathematical Sciences, University of Nottingham, University Park,
Nottingham NG7 2RD, United Kingdom}
\author{Andrzej Dragan}
\email{dragan@fuw.edu.pl}
\affiliation{Institute of Theoretical Physics, University of Warsaw, Pasteura 5, 02-093 Warsaw, Poland}
\author{Ivette Fuentes}
\email{ivette.fuentes@nottingham.ac.uk}
\affiliation{School of Mathematical Sciences, University of Nottingham, University Park,
Nottingham NG7 2RD, United Kingdom}
\affiliation{
Faculty of Physics, 
University of Vienna, 
Boltzmanngasse 5, 
A-1090 Vienna, 
Austria}

\date{\today}

\begin{abstract}
Moving cavities promise to be a suitable system for relativistic quantum information processing.  It has been shown that an inertial and a uniformly accelerated one-dimensional cavity can become entangled by letting an atom emit an excitation while it passes through the cavities, but the acceleration degrades the ability to generate entanglement. We show that in the two-dimensional case the entanglement generated is affected not only by the cavity's acceleration but also by its transverse dimension which plays the role of an effective mass.
\end{abstract}

%\keywords{}

\maketitle

\section{Introduction}

Entanglement is a quantum property which has a wide variety of applications in quantum information tasks such as teleportation and quantum cryptography.  In most canonical applications, two observers --- Alice and Bob --- each hold a subsystem and meet to prepare a bipartite maximally entangled state.  After the state has been prepared, they separate, taking with them their corresponding entangled subsystem which can be used to perform information processing tasks. Such protocols usually assume that spacetime is flat and that Alice and Bob move with non-relativistic speeds.  However, to impose more realistic conditions and gain additional insight into the properties of quantum correlations, we can analyze what differences arise when Alice and Rob (the relativistic Bob) move at relativistic speeds, accelerate, or are in the presence of a gravitational field. 

Early results in the field of relativistic quantum information suggested that entanglement of global modes is degraded from the perspective of observers moving in uniform acceleration \cite{fuentes-schuller2005,martinmartinez2009}. Such states are not useful to perform quantum information tasks as Alice and Rob must be able to store information in systems which they can manipulate. Therefore, they will require to store information in spatially localized states.  Moving cavities promise to be good candidates for this since quantum information can be encoded in cavity field modes which are localized within the cavities \cite{alsing2003}. It was shown that entanglement can be generated between two cavities, one inertial and one in uniform acceleration, by letting an atom interact with the modes of the cavities \cite{downes2011}. However, the result was obtained under the idealised assumptions of a massless one-dimensional system. Although useful as a proof of principle, an analysis of more realistic settings would shed light on the feasibility of entanglement generation in relativistic scenarios.

In this paper we consider the entanglement generated between the modes of moving two-dimensional bosonic cavities. In such a case, the transverse dimension of the cavity plays the role of an effective mass in the field equation. This has significant implications as the presence of mass (or effective mass) has an impact on the entanglement between cavity modes. For example, the degradation of entanglement between inertial and accelerated field modes is increased by several orders of magnitude when the fields are massive \cite{alphacentauri2012}, and the probability of the excitation of an atom moving through a cavity is lower for massive fields, which can be used to distinguish between inertial and non-inertial frames \cite{dragan2011}. We find that the entanglement generated by an atom interacting with the field of an inertial and an accelerated cavity is lower when the fields are massive, given some fixed cavity size. Since the transverse dimension contributes to the mass of the field, we can expect this type of degradation even in massless bosonic fields when they are considered in more realistic, two- and three-dimensional cases.

Throughout the paper we assume natural units ${\hbar=c=1}$.

\section{Physical Set-up}

We consider a pair of two-dimensional cavities that are in relative motion. The cavities carry a bosonic field which vanishes at perfectly reflecting mirrors which constitute the cavity walls. One of the cavities is uniformly accelerated in one direction only, which means that its other spatial dimension will remain inertial and hence unaffected by the motion. Following a scheme introduced by Browne and Plenio \cite{browne2003}, an atom is then passed through the two cavities thereby entangling the cavity field modes.

The stationary cavity can be described by the standard Minkowski spacetime coordinates $x^{\mu}=(t,x,y)$. We shall assume that this cavity, as described by an observer located at the origin of this coordinate system, has boundaries at $x_{\pm}$ and $y_{\pm}$ in the $x$ and $y$ dimensions, respectively. We denote the length of the cavity walls by $L^{i}\,=\,x^{i}_{+}-x^{i}_{-}$ where $i$ runs from $1$ to $2$ and denotes the spatial components of the coordinate $3$-vector. 

\begin{figure}[t!]
  \includegraphics[width=0.8\linewidth]{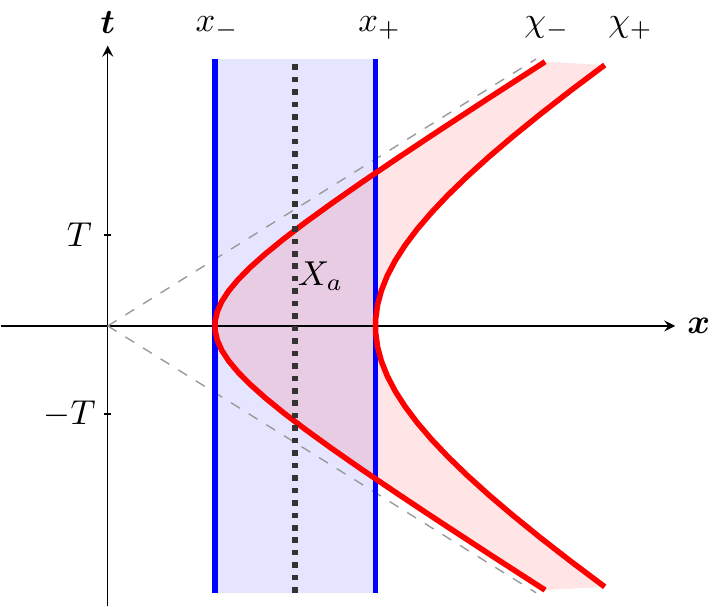}
  \caption{(Color online) Cavity set-up in the $x$ dimension. The cavities are positioned such that the accelerated (red) cavity becomes instantaneously aligned with the inertial (blue) cavity at $t=0$. The dotted line represents the path of the atom, which travels through the centre of the inertial cavity and comes into and out of alignment with the accelerated cavity at $t=\mp T$, crossing the centre of the cavity at $t=0$. The dashed lines denote Rindler horizons.}
  \label{fig:rinddiag}
\end{figure}

\begin{figure}[t!]
\includegraphics[width=0.8\linewidth]{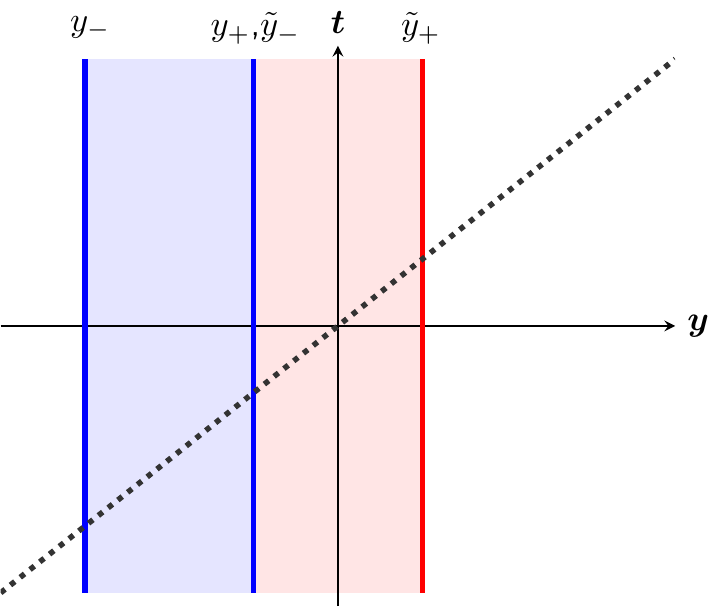}
\caption{(Color online) Cavity set-up in the $y$ dimension. We consider the cavities to be positioned side-by-side, such that the atom (dotted line), which is travelling through the inertial (blue) and then the accelerated (red) cavity at a constant velocity, crosses the centre of the accelerated cavity at $t=0$.}
\label{fig:ytdiagram}
\end{figure}

Next, we consider a uniformly accelerated cavity moving in the $x$ direction. The most useful coordinates to describe this motion are the Rindler coordinates $(\eta,\chi,y)$ defined via
\begin{equation}
\begin{aligned}
\label{MinkowskiTrans}
t &= \chi\sinh\left(\eta\right) \\
x &= \chi\cosh\left(\eta\right)
\end{aligned}
\end{equation}
and the $y$ coordinate is the same as the standard Minkowski $y$ coordinate. Analogously to the inertial cavity, we define the accelerating cavity walls by $\chi_{\pm}$ and $\tilde{y}_{\pm}$. Again, we denote the proper length of the cavity walls by $\tilde{L}^{i}\,=\,\chi^{i}_{+}-\chi^{i}_{-}$. The two cavity mirrors $\chi_{\pm}$ follow uniformly accelerated trajectories, and therefore accelerate with proper accelerations of $1/\chi_{\pm}$. We define $a$ to be the proper acceleration at the centre of Rob's cavity, such that $\chi_{\pm}=1/a\pm L/2$. In our scenario, we set all cavity lengths to be equal, i.e. $L^{i}=\tilde{L}_{i}=L$. Further, we choose $x_{\pm}=\chi_{\pm}$, $y_{-}=-3L/2$, $y_{+}=\tilde{y}_{-}=-L/2$ and $\tilde{y}_{+}=+L/2$. These coordinates mean that at the instant $t=\eta=0$, the two cavities are aligned, with their $x$ coordinates overlapping (see fig. \ref{fig:rinddiag}) and their $y$ coordinates positioned such that the cavities are side-by-side (fig. \ref{fig:ytdiagram}). We will take $L$ to be equal to unity for simplicity.

Finally, we consider the trajectory of the atom. We choose the atom to be always located at the centre of the inertial cavity in the $x$ dimension, while passing through $y=0$ at $t=0$ with constant velocity $v$ in the $y$ dimension. This can be written in $3$-vector notation as $x^{\mu}_{a}(t)=(t,X_{a},vt)$ where $X_{a}=(x_{+}-x_{-})/2$. For the dynamics of the fields, it will be more useful to parametrize the cavities and trajectories in terms of the atom's proper time. The relation between the proper time, which we denote as $\tau$, and the coordinate time $t$ is given by $t=\gamma\tau$ where $\gamma=1/\sqrt{1-v^2}$. The parametrization of the trajectory of the atom is then $x^{\mu}_{a}(\tau)=(\gamma\tau,X_{a},v\gamma\tau)$.

\section{Field dynamics}

Having examined the kinematics of the cavities, let us now describe the quantum fields that are carried by them. We consider the cavities to each carry a separate field, both of which vanish at cavity walls obeying Dirichlet boundary conditions. The fields are initially in the vacuum state according to the observers co-moving with each cavity. The Bogoliubov transformation of the field between Minkowski and Rindler reference frames is highly non-trivial \cite{bruschi2010}. It is well known that Alice's cavity, according to Rob, has some excitations and vice versa. Nonetheless, the initial state of the two cavities is separable regardless of the coordinates used to describe it.

The free dynamics of the fields are given by the massive Klein-Gordon equation, which takes the form
\begin{equation}
\frac{1}{\sqrt{-g}}\partial_{\mu}\left(g^{\mu\nu}\sqrt{-g}\partial_{\nu}\phi\right)-\kappa^{2}\phi=0.
\end{equation}
Here $g^{\mu\nu}$ is the inverse of the spacetime metric, $g=\det(g_{\mu\nu})$, and $\kappa$ is the bare mass of the field. In Alice's inertial frame the metric is given by $g^{(M)}_{\mu\nu}=\diag\left(-1,1,1\right)$ and the massive field equation becomes
\begin{equation}
-\partial_{tt}\phi +\left(\partial_{xx}+\partial_{yy}\right)\phi -\kappa^2\phi =0.
\end{equation}
In Alice's frame the mode inner product is defined as
\begin{equation}
\left(\phi_{1},\phi_{2}\right)_{M} = -i\int_{\Sigma_{M}} dxdy\left(\phi_{2}\partial_{t}\overline{\phi}_{1}-\overline{\phi}_{1}\partial_{t}\phi_{2}\right)
\end{equation}
where $\Sigma_{M}$ is a hypersurface of constant $t$. 
From here, we follow the usual quantisation procedure for fields by imposing the canonical commutation relation (CCR)
\begin{equation}
[\hat{a}_{mn},\hat{a}^{\dag}_{jk}]\,=\,\delta_{mj}\delta_{nk}
\end{equation}
where $\hat{a}_{mn}$ represents the quantum $(m,n)$-mode operator. This prescription also comes with the assumption of a state $|0\rangle_{A}$ for which $a_{mn}|0\rangle_{A}\,=\,0$ for all mode numbers $(m,n)$. For this particular choice of metric and coordinates, this state is called the \emph{Minkowski} vacuum~\cite{crispino2008}.

Using this quantisation method and imposing the field is real, we can expand the quantum field in terms of the classical field mode solutions and the quantum mode operators as,
\begin{equation}
\label{MinkowskiSolutions}
\hat{\phi}_{A}(t,x,y)=\sum_{n,m}N_{nm}u_{n}(x)u_{m}(y)e^{-i\omega_{nm}t}\hat{a}_{nm}+\text{h.c.}
\end{equation}
where
\begin{equation}
\begin{aligned}
u_{k}(x^{i})&=\sin\left[\frac{k\pi}{L_{i}}\left(x^{i}-x^{i}_{-}\right)\right]\\
N_{nm}&=\frac{\sqrt{2}}{\sqrt{\omega_{nm}L_{x}L_{y}}}\\
\omega_{nm}^{2}&=\left(\frac{n\pi}{L_{x}}\right)^{2}+\left(\frac{m\pi}{L_{y}}\right)^{2}+\kappa^{2}.
\end{aligned}
\end{equation}
We will therefore represent the quantum field contained within Alice's cavity via Eq.~\eqref{MinkowskiSolutions}. Analogously, the field contained within Rob's cavity can be quantised. In Rob's frame, described by Rindler coordinates, the metric is given by $g^{(R)}_{\mu\nu}=\diag\left(-\chi^{2},1,1\right)$ and so the Klein-Gordon equation becomes 
\begin{equation}
-\partial_{\eta\eta}\phi +\left(\chi\partial_{\chi}\chi\partial_{\chi}+\chi^{2}\partial_{yy}\right)\phi -\chi^{2}\kappa^2\phi =0
\end{equation}
with inner product
\begin{equation}
\left(\phi_{1},\phi_{2}\right)_{R}= -i\int_{\Sigma_{R}} d\chi dy\frac{1}{\chi}\left(\phi_{2}\partial_{\eta}\overline{\phi}_{1}-\overline{\phi}_{1}\partial_{\eta}\phi_{2}\right)
\end{equation}
and $\Sigma_{R}$ being a hypersurface of constant $\eta$. In this case, we impose the CCRs,
\begin{equation}
[\hat{b}_{mn},\hat{b}^{\dag}_{jk}]\,=\,\delta_{mj}\delta_{nk}.
\end{equation}
As for the Minkowski case, the Rindler coordinates induce a vacuum state which satisfies the property $\hat{b}_{mn}|0\rangle\,=\,0$ for all modes. We can finally write the quantum expansion of the accelerated cavities field as,
\begin{equation}
\label{RindlerSolutions}
\hat{\phi}_{R}(\eta,\chi,y)=\sum_{n,m}\tilde{N}_{nm}\tilde{u}_{nm}(\chi)\tilde{u}_{m}(y)e^{-i\tilde{\Omega}_{nm}\eta}\hat{b}_{nm}+\text{h.c.}
\end{equation}
Note that due to the trivial coordinate transformation in the $y$ dimension, $\tilde{u}_{m}(y)=u_{m}(y)$. The $\chi$ spatial function, on the other hand, is highly non-trivial and obeys the modified Bessel equation. 

The boundary conditions for the accelerated cavity walls are defined by $\chi_{\pm}$. This gives the mode solutions in the accelerated dimension as 
\begin{equation}
\begin{aligned}
\tilde{u}_{nm}(\chi)=&\,\Re\left[I_{i\tilde{\Omega}_{nm}}(\kappa_{m}\chi_{-})\right]K_{i\tilde{\Omega}_{nm}}(\kappa_{m}\chi)\\
&-K_{i\tilde{\Omega}_{nm}}(\kappa_{m}\chi_{-})\Re\left[I_{i\tilde{\Omega}_{nm}}(\kappa_{m}\chi)\right]\\
\kappa_{m}^{2}=&\left(\frac{m\pi}{\tilde{L}_{y}}\right)^{2}+\kappa^{2}
\end{aligned}
\end{equation}
where $I_{\alpha}(z)$ and $K_{\alpha}(z)$ are the modified Bessel functions of the first and second kind, respectively. The quantities $\tilde{N}_{nm}$ and $\tilde{\Omega}_{nm}$ are functions of the mode numbers $(n,m)$, the acceleration of Rob's cavity and the bare mass of the field. They are only analytically closed functions for the massless $(1+1)$-dimensional case of \cite{downes2011}. We shall evaluate them numerically for a specified acceleration and bare mass. Note that we now have a mode-dependent mass term $\kappa_{m}$, hence the field gains an effective mass due to the presence of the transverse spatial dimension.

\section{Atom-field interaction}

We describe the atom as a two-level system with a ground state $\ket{g}$ and an excited state $\ket{e}$. The cavity modes can become entangled by their interaction with the atom, which passes through the cavities.  To achieve this, the atom is initially prepared in its exited state, and sent through the cavities with a non-zero probability of emitting an excitation in either of them. The state of the atom is subsequently measured. If the atom is found to be in the ground state, an interaction has occurred; however, it is impossible to discriminate which cavity field has been excited by the atom without further measurements. The final state of the system must therefore be a superposition of both possibilities, resulting in an entangled state of the two cavities \cite{browne2003}. Throughout this article we will describe all interactions in the interaction picture, which allows us to compute the dynamics of a state by considering only the interaction terms of a system Hamiltonian. For a discussion of the interaction picture and its details we refer the interested reader to~\cite{brown2013,bruschi2013,greiner1996}.

We model the atom with the simple case of an Unruh-DeWitt detector, described by a characteristic frequency $\Delta$ along with raising and lowering operators $\hat{d}^{\dagger}$ and $\hat{d}$, respectively. The quantum mechanical description of the atom is given by its monopole operator
\begin{equation}
\hat{M}(\tau)=\hat{d}^\dagger e^{i\Delta\tau}+\hat{d}e^{-i\Delta\tau}.
\end{equation}
The interaction Hamiltonians are given by  
\begin{equation}
\label{UnruhDeWittHamiltonian}
\hat{H}_{A/R}^{I}(\tau)=\epsilon_{A/R}(\tau)\hat{M}(\tau)\hat{\phi}_{A/R}(\tau)
\end{equation}
where $\hat{\phi}_{A}$ and $\hat{\phi}_{R}$ are field operators given by eqs.~\eqref{MinkowskiSolutions} and~\eqref{RindlerSolutions} which are evaluated along the world line of the atom with proper time $\tau$. The switching functions $\epsilon_A$ and $\epsilon_R$ represent the strength of interaction and in general can be time-dependent. We choose them both to be sine functions of the detector's proper time of the form
\begin{equation}
\epsilon_{A/R}(\tau) = \epsilon(\tau)=\epsilon\sin^{2}\left(2\pi v\gamma\tau\right),
\end{equation}
which spread the interaction smoothly over the whole interaction interval and vanish at the cavity boundaries, representing the atom going out of alignment with the cavities.

The evolution of the entire system can be written as $\ket{\psi}=U_{R}U_{A}\ket{\psi(0)}$ where $\ket{\psi(0)}$ is the initial state of the system, and $U_{A}$ and $U_{R}$ are the unitary operators that evolve Alice and Rob's subsystems. This corresponds to the state first evolving under the interaction of Alice's Hamiltonian, followed by Rob's. Assuming that the cavities are initially in the vacuum state and the atom is excited, i.e. $\ket{\psi(0)}=\ket{0}_{A}\ket{0}_{R}\ket{e}$, the evolution of the state to first order in perturbation theory becomes
\begin{equation}
\ket{\psi}=\left(\mathrm{id}-i\int d\tau \left(H_{A}^{I}(\tau)+H_{R}^{I}(\tau)\right)\right)\ket{0}_{A}\ket{0}_{R}\ket{e}
\end{equation}
where the integration is over the interaction time of the atom and the cavities and $\mathrm{id}$ represents the identity operator. The kinematic set-up of the cavities gives the interaction time interval for Alice's and Rob's cavities as $\tau\in\left[-3T, -T\right]$ and $\tau\in\left[-T,T\right]$, respectively, where we can now write $T = 1/2v\gamma$.

In our scheme we will be interested in events where the atom has been detected in the ground state after passing through the cavities. Denoting the post-selected state of the system as $\ket{\phi}$, we can write it as

\begin{widetext}
\begin{equation}
\ket{\phi} = -i\!\int\!\!\mathrm{d}\tau\bra{g}\hat{d}e^{-i\Delta\tau}\ket{e}M(\tau)\epsilon(\tau)\left[\hat{\phi}_{A}(\tau)+\hat{\phi}_{R}(\tau)\right]\ket{0}_{A}\ket{0}_{R}.
\end{equation}
\end{widetext}
We observe the only non-zero contributions to this state come from the $\hat{d}\hat{a}^{\dag}$ and $\hat{d}\hat{b}^{\dag}$ terms of the Hamiltonians \eqref{UnruhDeWittHamiltonian}. Expanding the post-selected state as a superposition over single particle states, we obtain
\begin{equation}
  \ket{\phi}=\displaystyle\sum_{n,m}\left[F_{nm}^{A}\hat{a}^{\dag}_{nm}+F_{nm}^{R}\hat{b}^{\dag}_{nm}\right]\ket{0}_{A}\ket{0}_{R}
\end{equation}
with
\begin{equation}
\begin{aligned}
F_{nm}^{A}=&-i\sin(n\pi/2)\int\limits_{-3T}^{-T}d\tau\Lambda(\tau) e^{+i\omega_{nm}\gamma\tau}\\
F_{nm}^{R}=&-i\int\limits_{-T}^{T}d\tau\Lambda(\tau)\tilde{u}_{nm}(\chi(\tau))e^{+i\tilde{\Omega}_{nm} \arctanh(a\gamma\tau)}
\end{aligned}
\end{equation}
where $\chi(\tau)=\sqrt{1/a^{2}-\gamma^{2}\tau^{2}}$, $T=1/2v\gamma$ and we have denoted
\begin{equation}
\Lambda(\tau)=\epsilon(\tau)\sin\left(m\pi\left(v\gamma\tau-1/2\right)\right)e^{-i\Delta\tau}.
\end{equation}
It should be noted that for a given atom velocity $v$, the proper acceleration at the centre of Rob's cavity has a maximum value. Since the term under the square root in $\chi(\tau)$ has to be positive, we can write $a\le (\gamma|\tau|)^{-1}$ for all  $\tau\in[-T,T]$. As this inequality has to be true for all allowed $\tau$, we can establish an upper bound on the cavity's acceleration by maximising $|\tau|$. Substituting $|\tau|_{\mathrm{max}}\,=\,T\,=\,(2v\gamma)^{-1}$ into the previous inequality gives us an upper bound on acceleration as $a\le2v$. Physically, this means the walls of Rob's cavity cannot crash into the atom. The maximum acceleration at the centre of Rob's cavity is therefore $a=2$ in the limit $v\to 1$, noting that in this limit one side of Rob's cavity approaches the Rindler horizon and therefore its proper acceleration diverges.

\section{Entanglement Calculation}

Let us determine the degree of entanglement shared by the cavities after the atom has passed through the cavities. The state of the cavities is pure, since we only take into account the post-selected events when the atom is found in the ground state after the interaction, therefore we can use the Von Neumann entropy as a valid measure of non-local correlations. We first find the reduced density matrix $\hat{\rho}_{R}=\text{tr}_{A}(\ket{\phi}\bra{\phi})$ of Robs's cavity. Since $_A\!\bra{0}\hat{a}^{\dag}_{nm}\ket{0}_{A}=\,_A\!\bra{0}\hat{a}_{nm}\ket{0}_{A}=0$ and $_A\!\bra{0}\hat{a}_{nm}\hat{a}^{\dag}_{ij}\ket{0}_{A}=\delta_{ni}\delta_{mj}$, we have
\begin{equation}
\label{densitymatrix}
\hat{\rho}_{R}=\sum_{n,m}|F_{nm}^{A}|^{2} \ket{0}_{R}\!\bra{0}+\sum_{nmij}F_{nm}^{R}\,\bar{F}_{ij}^{R}\hat{b}^{\dag}_{nm}\ket{0}_{R}\!\bra{0}\hat{b}_{ij}.
\end{equation}
We find numerically that there always exists a point where successive terms stop contributing non-trivially to the sum. We can therefore truncate the matrix involved in the expression, and instead consider a finite-dimensional matrix. We arrange the mode integrals into a vector as 
\begin{equation}
\vec{F}=\left(F_{11}^{R},\ldots, F_{1N}^{R},\ldots, F_{N1}^{R},\ldots, F_{NN}^{R}\right).
\end{equation}
From this vector we can construct a matrix representation of Rob's state in the basis $\{\hat{b}^{\dag}_{nm}\ket{0}_{R}\}$ as $\rho_{R}:=\left(\sum_{n,m}|F_{nm}^{A}|^{2}\right)\oplus\left(\vec{F}\otimes\vec{F}^{\dagger}\right)$. Our newly constructed matrix then takes the form
\begin{equation}
 \rho_{R} =
 \begin{pmatrix}
  \displaystyle\sum_{n,m}|F_{nm}^{A}|^{2} & 0 & \cdots & 0 \\
  0 & |F_{11}^{R}|^{2} & \cdots & F_{11}^{R}\overline{F}_{NN}^{R} \\
  \vdots  & \vdots  & \ddots & \vdots  \\
  0 & F_{NN}^{R}\overline{F}_{11}^{R} & \cdots & |F_{NN}^{R}|^{2}
 \end{pmatrix}
\end{equation}
for a numerically established truncation point $N$. Here we have defined the first element of the matrix to be the coefficient of $\ket{0}_{R}\bra{0}$. After the renormalization $\hat{\rho}_{R}\rightarrow\hat{\rho}_{R}/\tr(\hat{\rho}_{R})$ where $\text{tr}(\hat{\rho}_{R})=\sum_{n,m}\left[|F_{nm}^{A}|^{2}+|F_{nm}^{R}|^{2}\right]$, the state can be used to evaluate the Von Neumann entropy $S(\rho)=-\tr\left(\rho\log\rho\right)$. One then finds the eigenvalues of the truncated matrix and computes the entropy via $S(\rho_{R})=-\sum_{k}\lambda_{k}\log(\lambda_{k})$ where $\lambda_{k}$ are the eigenvalues of $\rho_{R}$.
\begin{figure}[t]
  \includegraphics[width=8cm]{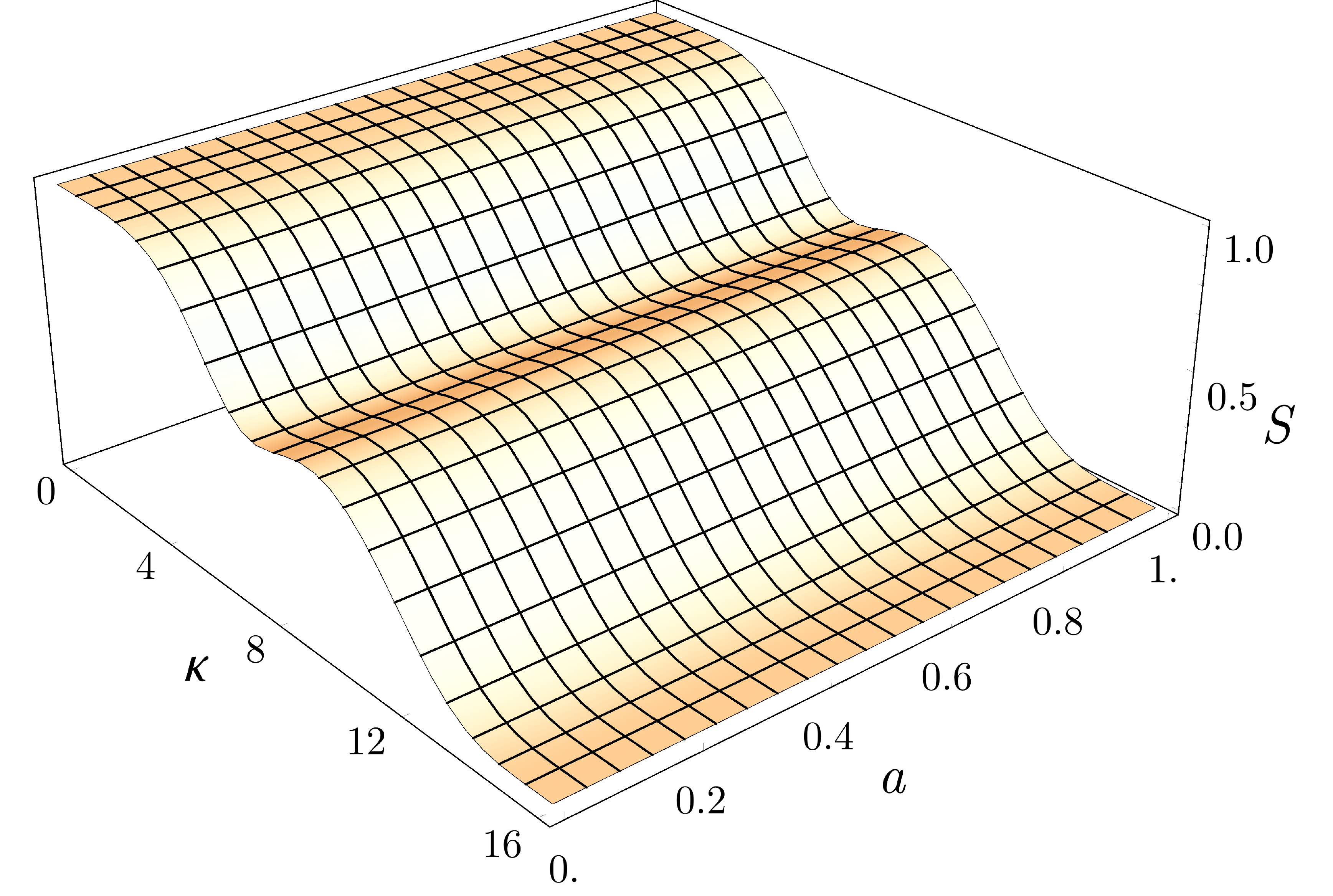}
  \caption{(Color online) Entanglement generated between the two cavities as a function of acceleration at the centre of Rob's cavity, $a$, and the bare mass of his field $\kappa$. Here we set all cavity lengths to unity, $\Delta=\sqrt{2\pi^{2}}$ and $v=1/2$.}
  \label{big2dplot}
\end{figure}
We find the entanglement generated between the cavities decreases monotonically as a function of the proper acceleration at the centre of Rob's cavity. One can see in fig.~\ref{big2dplot} that the effect of acceleration is much smaller when compared with the effect due to mass. As previously mentioned, the kinematic set-up of the cavities dictates the maximum acceleration of Rob's cavity. To access higher accelerations, and hence generate smaller amounts of entanglement, the velocity of the two-level system must be increased.

Additionally, we find that the entanglement degradation oscillates as a function of the bare mass of Rob's field. To understand the oscillations in the plot, we can consider the case of zero acceleration, i.e. $a=0$. The integrals $F_{nm}^{R}$ can then be explicitly calculated. The resulting expression takes the form
\begin{equation}
F_{nm}^{R}=f_{nm}(\kappa)\left(1-(-1)^{n}e^{i g_{nm}(\kappa)}\right)
\end{equation}
where
\begin{equation}
g_{nm}(\kappa)=\frac{1}{v}\left(\Delta\sqrt{1-v^{2}}-\sqrt{\pi^{2}n^{2}+\pi^{2}m^{2}+\kappa^{2}} \right)
\end{equation}
and $f_{nm}(\kappa)$ is a polynomial in $\kappa$ of order $\mathcal{O}(\kappa^{-6})$. There are points of constructive resonance where $1-(-1)^{n}e^{i g_{nm}(\kappa)}=2$, and it is these resonances that contribute to the local maxima observed in the entanglement between the cavities. Physically, this corresponds to a resonance between the internal energy gap of the detector and the energy of the quanta contained within Rob's cavity.

The results show that the introduction of field mass degrades our ability to entangle the two cavities. However, the entanglement degradation can be in principle avoided by adjusting the physical characteristics of the cavities~\cite{downes2011}.

\section{Conclusions}
We have examined the entanglement generated between the modes of two cavities in relative motion when interacting with a two-level system. We showed that the physical set-up is robust against the effects of acceleration, which could be exploited to generate entangled states between an inertial and an accelerated cavity for use in quantum information protocols.  Moreover, we found that the mass of the field contained within the accelerated cavity reduces our ability to generate entangled states, introducing a damped undulatory degradation. The extra spatial dimensions contribute to the mass of the field and, therefore, we can expect the degradation effect to occur in experimental implementations that use massless bosonic fields. In the case of optomechanical experiments, the current maximum acceleration achievable is of the order $20g$~\cite{ursin}. Assuming this maximum acceleration is obtained at the $\chi_{-}$ boundary, we find that the acceleration at the centre of Rob's cavity is (with full units restored) $a\,=\,20g/(1+10gL/c^{2})\approx 20g$ for any realistic choice of $L$. In our analysis we assumed dimensionless acceleration was up to the order unity and hence the physical acceleration is of the order $c^{2}\gg20g$. Therefore, for optomechanical settings, the detrimental effects of acceleration will be minimal and the most significant degradation will occur due to mass (bare, effective or both). 

Another possible use of our analysis could be in the ability to distinguish between inertial and accelerated observers. Given that extra spatial dimensions act as an effective mass for the bosonic field, one could distinguish inertial frames from non-inertial frames following the setting of Dragan et al.~\cite{dragan2011}.

Future directions of work could be to extend the model of scalar fields to Dirac fields and to use the detectors themselves for entanglement extraction.

We would like to thank Gerardo Adesso, Nicolai Friis, Sara Tavares and Jorma Louko for interesting discussions and useful comments. A.~R.~L. was supported by the EPSRC Doctoral Prize. A.~D. thanks for the financial support to the National Science Center, Sonata BIS Grant No. 2012/07/E/ST2/01402. I.~F. thanks EPSRC [CAF Grant EP/G00496X/2] for financial support. 

\bibliography{main}

\end{document}